\documentclass[letterpaper,12pt,notoc]{JHEP3}
\def\pd{\partial}
\def\mc{\mathcal}

\usepackage{graphicx}
\usepackage{amsmath}
\usepackage{amssymb}
\usepackage{caption}
\usepackage{subcaption}
\preprint{ \hbox{}\hfill arXiv: 1411.4542}

\title{Noncompact gauging of N=2 7D supergravity and AdS/CFT holography}
\author{Parinya Karndumri\\
String Theory and Supergravity Group, Department
of Physics, Faculty of Science, Chulalongkorn University, 254 Phayathai Road, Pathumwan, Bangkok 10330, Thailand\\
E-mail: \email{parinya.ka@hotmail.com}}

\abstract{Half-maximal gauged supergravity in seven dimensions coupled to $n$
vector multiplets contains $n+3$ vectors and $3n+1$ scalars parametrized by $\mathbb{R}^+\times SO(3,n)/SO(3)\times SO(n)$ coset
manifold. The two-form field in the gravity multiplet can be
dualized to a three-form field which admits a topological mass term.
Possible non-compact gauge groups take the form of $G_0\times
H\subset SO(3,n)$ with a compact group $H$. $G_0$ is one of the five
possibilities; $SO(3,1)$, $SL(3,\mathbb{R})$, $SO(2,2)$, $SO(2,1)$
and $SO(2,2)\times SO(2,1)$. We investigate all of these possible
non-compact gauge groups and classify their vacua. Unlike the gauged
supergravity without a topological mass term, there are new
supersymmetric $AdS_7$ vacua in the $SO(3,1)$ and $SL(3,\mathbb{R})$
gaugings. These correspond to new $N=(1,0)$ superconformal field
theories (SCFT) in six dimensions. Additionally, we find a class of
$AdS_5\times S^2$ and $AdS_5\times H^2$ backgrounds with $SO(2)$ and
$SO(2)\times SO(2)$ symmetries. These should correspond to $N=1$
SCFTs in four dimensions obtained from twisted compactifications of
six-dimensional field theories on $S^2$ or $H^2$. We also study RG
flows from six-dimensional $N=(1,0)$ SCFT to $N=1$ SCFT in four
dimensions and RG flows from a four-dimensional $N=1$ SCFT to a
six-dimensional SYM in the IR. The former are driven by a vacuum
expectation value of a dimension-four operator dual to the
supergravity dilaton while the latter are driven by vacuum
expectation values of marginal operators.}

\keywords{AdS-CFT correspondence, Gauge/Gravity Correspondence and
Supergravity Models}
\begin{document}
\section{Introduction}
Gauged supergravities play an important role in string/M theory
compactification and gauge/gravity correspondence. Generally, a
gauge supergravity theory admits many types of gauge groups namely
compact, non-compact and non-semisimple groups, and different types
of gauge groups give rise to different vacuum structures. Gauged
supergravity theories may be accordingly classified into two
categories by the vacua they admit. AdS supergravities are theories
admitting a maximally supersymmetric AdS space as a vacuum solution
while those with a half-maximally supersymmetric domain wall vacuum
are called domain-wall supergravities. The former is useful in the
context of the AdS/CFT correspondence \cite{maldacena}, and the
latter is relevant in the DW/QFT correspondence
\cite{DW/QFT_townsend,correlator_DW/QFT}.
\\
\indent  The study of $N=(1,0)$ superconformal field
theories (SCFT) in the context of AdS$_7$/CFT$_6$ correspondence has originally done by orbifolding the $AdS_7\times S^4$ geometry of M-theory giving rise to the gravity dual of $N=(2,0)$ SCFT \cite{Berkooz_6D_dual,AdS7_orbifold1,AdS7_orbifold2}. And, recently, many $AdS_7$ solutions to type IIA string theory have been identified in \cite{All_AdS7}. These backgrounds are dual
to $N=(1,0)$ SCFTs in six dimensions, and the holographic study of these
SCFTs has been given in \cite{Dual_of_6DN10}. Furthermore, a number of $N=(1,0)$ SCFTs in six dimensions
have been found and classified in the context of F-theory in \cite{6DSCFT_from_Ftheory}.  It would be desirable to have a description of these SCFT in terms of the gravity solutions to seven-dimensional gauged supergravity. However, it has been pointed out in \cite{Van_Riet} that $AdS_7$ solutions found in \cite{All_AdS7} cannot be obtained from seven-dimensional gauged supergravity.
\\
\indent In the framework of
seven-dimensional gauged supergravity, there are only a few results in the holography of $N=(1,0)$ SCFTs. It has been proposed in \cite{Ferrara_AdS7CFT6} that the $N=(1,0)$ SCFTs arising in the M5-brane world-volume theories should be described by $N=2$ seven-dimensional gauged supergravity and its matter-coupled version. A non-supersymmetric holographic RG flow within pure $N=2$ gauged supergravity has been studied in \cite{non_SUSY7Dflow}, and recently, new supersymmetric $AdS_7$
critical points and holographic RG flows between these critical
points have been explored in \cite{7D_flow}. The gauged supergravity considered
in \cite{7D_flow} is the $N=2$ gauged supergravity coupled to three
vector multiplets resulting in $SO(4)\sim SU(2)\times SU(2)$ gauge group with two coupling constants for the two $SU(2)$'s. When these couplings are equal, the theory can be embedded in eleven dimensions by using the reduction ansatz recently obtained in \cite{SO4_7Dfrom11D}.
\\
\indent To find more supersymmetric $AdS_7$ backgrounds, in this paper, we will consider the $N=2$ gauged
supergravity in seven dimensions coupled to a number of vector
multiplets with non-compact gauge groups. The gauged supergravity is
obtained from coupling pure $N=2$ supergravity constructed in
\cite{Pure_N2_7D1} to vector multiplets \cite{Eric_N2_7D}.
Furthermore, the two-form field in the supergravity multiplet can be
dualized to a three-form field \cite{Park_7D}. It turns out to be possible to add a
topological mass term to this three-form field resulting in a gauged
supergravity with a massive three-form field \cite{Eric_N2_7Dmassive}.
The latter differs considerably from the theory without topological
mass in the sense that it is possible to have maximally
supersymmetric $AdS_7$ backgrounds.
\\
\indent We will see that there are new $AdS_7$ critical points for
non-compact gauging of the $N=2$ supergravity with topological mass
term. These provide more examples of $AdS_7$ solutions with sixteen
supercharges. We will also find that some non-compact gauge groups
admit $AdS_5\times S^2$ and $AdS_5\times H^2$ geometries as a
background solution. In the context of twisted field theories, these
solutions should describe a six-dimensional SCFT wrapped on a
two-dimensional Riemann surface. In the IR, the six-dimensional SCFT would flow to
another SCFT in four dimensions. These results give new $AdS_5$
backgrounds dual to $N=1$ four-dimensional SCFTs.
\\
\indent The holographic study of twisted field theories has originally been applied to $N=4$ SYM \cite{MN_nogo}. Until now, the method has been applied to other dimensions, see for example \cite{Carlos_F4_flow,3DSUGRA_fromD3,Benini_Bobev2,Bobev_4DSCFT}. In \cite{Bobev_4DSCFT}, $AdS_5$ solutions from a truncation of the maximal $N=4$ gauged supergravity in seven dimensions have been found. These $AdS_5$ geometries correspond to a class of $N=1$ SCFTs in four dimensions obtained from M5-branes wrapped on complex curves. In this paper, we will give more examples of these $N=1$ SCFTs by finding new $AdS_5$ geometries with eight supercharges in the half-maximal $N=2$ gauged supergravity. We also give some examples of RG flows from six-dimensional SCFTs to these four-dimensional SCFTs. Furthermore, we find an RG flow from a four-dimensional $N=1$ SCFT in the UV to a six-dimensional $N=(1,0)$ SYM in the IR. This flow gives another example of the flows considered in \cite{4D_5D_flow} in which the flows from $N=4$ SYM to six-dimensional $N=(2,0)$ SCFT and $N=2^*$ theory to five dimensional $N=2$ SCFT have been studied.
\\
\indent The paper is organized as follow. In section
\ref{7D_gaugedN2}, we describe $N=2$ gauged supergravity in seven
dimensions to set up the notation and discuss all possible
non-compact gauge groups. These gauge groups will be studied in
detail in section \ref{SO3_1}, \ref{SL3}, \ref{SO2_2} and
\ref{SO2_2_1} in which possible vacua and RG flow solutions will be
given. In section \ref{conclusion}, we give a summary of the results
and some conclusions.

\section{Seven-dimensional $N=2$ gauged supergravity coupled to $n$ vector multiplets}\label{7D_gaugedN2}
In this section, we give a description of the matter-coupled minimal
$N=2$ gauged supergravity in seven dimensions with topological mass
term. All of the notations are the same as those in
\cite{Eric_N2_7Dmassive} to which the reader is referred to for
further details.
\\
\indent A general matter-coupled theory is constructed by coupling $n$ vector multiplets to pure $N=2$ supergravity constructed in \cite{Pure_N2_7D1}. The supergravity multiplet $(e^m_\mu, \psi^A_\mu,
A^i_\mu,\chi^A,B_{\mu\nu},\sigma)$ consists of the graviton, two gravitini, three vectors, two spin-$\frac{1}{2}$ fields, a two-form field and a real scalar, the dilaton. The only matter mutiplet is the vector multiplet $(A_\mu,\lambda^A,\phi^i)$ consisting of a vector field, two gauginos and three scalars. We use the convention that curved and flat space-time indices are denoted by $\mu,\nu,\ldots$ and
$m,n,\ldots$, respectively. Spinor fields, $\psi^A_\mu$, $\chi^A$, $\lambda^A$, and the supersymmetry parameter $\epsilon^A$ are symplectic-Majorana spinors transforming as doublets of the R-symmetry $USp(2)_R\sim SU(2)_R$. From now on, the $SU(2)_R$ doublet indices $A,B=1,2$ will be dropped. Indices $i,j=1,2,3$ label
triplets of $SU(2)_R$.
\\
\indent The supergravity theory coupled to $n$ vector multiplets has
$SO(3,n)$ global symmetry. The $n$ vector multiplets will be
labelled by an index $r=1,\ldots n$. There are then $n+3$ vector
fields in total. Accordingly, only a subgroup $G$ of the global
symmetry $SO(3,n)$ of dimension $\textrm{dim}\, G\leq n+3$ can be
gauged. Possible gauge groups with structure constants
$f_{IJ}^{\phantom{IJ}K}$ and gauge algebra
\begin{equation}
\left[T_I,T_J\right]=f_{IJ}^{\phantom{IJ}K}T_K
\end{equation}
can be gauged provided that the $SO(3,n)$ Killing form $\eta_{IJ}$, $I,J=1,\ldots n+3$, is invariant under $G$
\begin{equation}
f_{IK}^{\phantom{IJ}L}\eta_{LJ}+f_{JK}^{\phantom{IJ}L}\eta_{LI}=0\, .\label{consistency}
\end{equation}
Since $\eta_{IJ}$ has only three negative eigenvalues, any gauge
group can have three or less compact generators or three or less
non-compact generators. It follows from \eqref{consistency} that the part of $\eta_{IJ}$ corresponding to each simple subgroup $G_\alpha$ of $G$ must be a multiple of the $G_\alpha$ Killing form. Therefore, possible non-compact gauge groups take the form of $G_0\times H$ with a compact group $H\subset
SO(3,n)$ of dimension $\textrm{dim}\, H\leq (n+3-\textrm{dim}\,
G_0)$ \cite{Eric_N2_7Dmassive}. The $G_0$ factor can only be one of the five possibilities:
$SO(3,1)$, $SL(3,\mathbb{R})$, $SO(2,1)$, $SO(2,2)\sim SO(2,1)\times
SO(2,1)$ and $SO(2,2)\times SO(2,1)$.
\\
\indent Apart from the dilaton $\sigma$ which is a singlet under the
gauge group, there are $3n$ scalar fields $\phi^{ir}$ parametrized
by $SO(3,n)/SO(3)\times SO(n)$ coset manifold. The associated coset
representative $L=(L_I^{\phantom{I}i},L_I^{\phantom{I}r})$
transforms under the global $SO(3,n)$ and the local $SO(3)\times
SO(n)$ by left and right multiplications, respectively. Its inverse
is denoted by $L^{-1}=(L^I_{\phantom{s}i},L^I_{\phantom{s}r})$ with
the relations $L^I_{\phantom{s}i}=\eta^{IJ}L_{Ji}$ and
$L^I_{\phantom{s}r}=\eta^{IJ}L_{Jr}$.
\\
\indent The two-form field $B_{\mu\nu}$ can be dualized to a
three-form field $C_{\mu\nu\rho}$ which admits a topological mass
term
\begin{equation}
\frac{h}{36}\epsilon^{\mu_1\ldots\mu_7}H_{\mu_1\ldots \mu_4}C_{\mu_5\ldots\mu_7}
\end{equation}
where the four-form field strength is defined by $H_{\mu\nu\rho\sigma}=4\pd_{[\mu}C_{\nu\rho\sigma]}$.
\\
\indent The bosonic Lagrangian of the $N=2$ massive-gauged supergravity is then given by
\begin{eqnarray}
e^{-1}\mc{L}&=&\frac{1}{2}R-\frac{1}{4}e^\sigma a_{IJ}F^I_{\mu\nu}F^{J\mu\nu}
-\frac{1}{48}e^{-2\sigma}H_{\mu\nu\rho\sigma}H^{\mu\nu\rho\sigma}-\frac{5}{8}\pd_\mu\sigma \pd^\mu\sigma
-\frac{1}{2}P^{ ir}_\mu P^\mu_{ir}\nonumber \\
& &-\frac{1}{144\sqrt{2}}e^{-1}\epsilon^{\mu_1\ldots
\mu_7}H_{\mu_1\ldots\mu_4}
\omega_{\mu_5\ldots\mu_7}+\frac{1}{36}he^{-1}\epsilon^{\mu_1\ldots
\mu_7}H_{\mu_1\ldots\mu_4} C_{\mu_5\ldots\mu_7}-V\nonumber \\
 \label{7Daction}
\end{eqnarray}
where the scalar potential is given by
\begin{equation}
V=\frac{1}{4}e^{-\sigma}\left(C^{ir}C_{ir}-\frac{1}{9}C^2\right)+16h^2e^{4\sigma}
-\frac{4\sqrt{2}}{3}he^{\frac{3\sigma}{2}}C\,.
\end{equation}
The Chern-Simons term is defined by
\begin{equation}
\omega_{\mu\nu\rho}=3\eta_{IJ}F^I_{[\mu\nu}A^J_{\rho]}-f_{IJ}^{\phantom{sa}K}A^I_{\mu}\wedge
A^J_{\nu}\wedge A_{\rho K}
\end{equation}
with
$F^I_{\mu\nu}=2\pd_{[\mu}A^I_{\nu]}+f_{JK}^{\phantom{sas}I}A^J_{\mu}A^K_{\nu}$.
\\
\indent We are going to find supersymmetric bosonic background solutions, so the supersymmetry transformations of fermions are needed. Since, in the following analysis, we will set $C_{\mu\nu\rho}=0$, we will accordingly give the supersymmetry transformations with all fermions and the three-form field vanishing. These are given by
\begin{eqnarray}
\delta \psi_\mu &=&2D_\mu
\epsilon-\frac{\sqrt{2}}{30}e^{-\frac{\sigma}{2}}C\gamma_\mu
\epsilon-\frac{i}{20}e^{\frac{\sigma}{2}}F^i_{\rho\sigma}\sigma^i\left(3\gamma_\mu
\gamma^{\rho\sigma}-5\gamma^{\rho\sigma}\gamma_\mu\right)\epsilon
-\frac{4}{5}he^{2\sigma}\gamma_\mu \epsilon,\label{delta_psi}\\
\delta \chi &=&-\frac{1}{2}\gamma^\mu\pd_\mu \sigma
\epsilon-\frac{i}{10}e^{\frac{\sigma}{2}}F^i_{\mu\nu}\sigma^i\gamma^{\mu\nu}\epsilon
+\frac{\sqrt{2}}{30}e^{-\frac{\sigma}{2}}C\epsilon-\frac{16}{5}e^{2\sigma}h\epsilon,\label{delta_chi}\\
\delta \lambda^r &=&-i\gamma^\mu
P^{ir}_\mu\sigma^i\epsilon-\frac{1}{2}e^{\frac{\sigma}{2}}F^r_{\mu\nu}\gamma^{\mu\nu}\epsilon
-\frac{i}{\sqrt{2}}e^{-\frac{\sigma}{2}}C^{ir}\sigma^i\epsilon\, .\label{delta_lambda}
\end{eqnarray}
The covariant derivative of $\epsilon$ is defined by
\begin{equation}
D_\mu\epsilon=\pd_\mu
\epsilon+\frac{1}{4}\omega_{\mu}^{ab}\gamma_{ab}+\frac{i}{4}\sigma^i\epsilon^{ijk}Q_{\mu
jk}
\end{equation}
where $\gamma^a$ are space-time gamma matrices.
\\
\indent The quantities appearing in the Lagrangian and the
supersymmetry transformations are defined by
\begin{eqnarray}
P_\mu^{ir}&=&L^{Ir}\left(\delta^K_I\pd_\mu+f_{IJ}^{\phantom{sad}K}A_\mu^J\right)L^i_{\phantom{s}K},
\qquad
Q^{ij}_\mu=L^{Ij}\left(\delta^K_I\pd_\mu+f_{IJ}^{\phantom{sad}K}A_\mu^J\right)L^i_{\phantom{s}K},\nonumber
\\
C_{ir}&=&\frac{1}{\sqrt{2}}f_{IJ}^{\phantom{sad}K}L^I_{\phantom{s}j}L^J_{\phantom{s}k}L_{Kr}\epsilon^{ijk},
\qquad
C=-\frac{1}{\sqrt{2}}f_{IJ}^{\phantom{sad}K}L^I_{\phantom{s}i}L^J_{\phantom{s}j}L_{Kk}\epsilon^{ijk},\nonumber \\
C_{rsi}&=&f_{IJ}^{\phantom{sad}K}L^I_{\phantom{s}r}L^J_{\phantom{s}s}L_{Ki},\qquad
a_{IJ}=L^i_{\phantom{s}I}L_{iJ}+L^r_{\phantom{s}I}L_{rJ},\nonumber \\
F^i_{\mu\nu}&=&L^{\phantom{I}i}_{I}F^I,\qquad
F^r_{\mu\nu}=L^{\phantom{I}r}_{I}F^I\, .
\end{eqnarray}
\indent In the following sections, we will study all possible
non-compact gauge groups $G_0$ without the compact $H$ factor. This
is a consistent truncation since all scalar fields we retain are $H$
singlets. All of the solutions found here are automatically solutions
of the gauged supergravity with $G_0\times H$ gauge group according
to the result of Schur's lemma as originally discussed in
\cite{warner}.
\\
\indent Before going to the computation, we will give a general parametrization of the $SO(3,n)/SO(3)\times SO(n)$ coset. We first introduce $(n+3)^2$ basis elements of a general $(n+3)\times (n+3)$ matrix as follow
\begin{equation}
(e_{IJ})_{KL}=\delta_{IK}\delta_{JL}\, .
\end{equation}
The composite $SO(3)\times SO(n)$ generators are given by
\begin{eqnarray}
SO(3)&:&\qquad J^{(1)}_{ij}=e_{ji}-e_{ij},\qquad i,j=1,2,3,\nonumber \\
SO(n)&:&\qquad J^{(2)}_{rs}=e_{s+3,r+3}-e_{r+3,s+3},\qquad
r,s=1,\ldots, n\, .
\end{eqnarray}
The non-compact generators corresponding to the $3n$ scalars are given by
\begin{equation}
Y^{ir}=e_{i,r+3}+e_{r+3,i}\, .
\end{equation}
The coset representative in each case will be given by an exponential of the relevant $Y^{ir}$ generators.

\section{$SO(3,1)$ gauge group}\label{SO3_1}
The minimal scalar coset for embedding $SO(3,1)$ gauge group is $SO(3,3)/SO(3)\times SO(3)$. We will choose the gauge structure constants to be
\begin{equation}
f_{IJK}=-g(\epsilon_{ijk},\epsilon_{rsi}),\qquad i,j,r,s=1,2,3
\end{equation}
from which we find $f_{IJ}^{\phantom{ssa}K}=\eta^{KL}f_{IJL}$ with
$\eta^{IJ}=(-1,-1,-1,1,1,1)$. Together with the dilaton $\sigma$,
there are ten scalars in this case. At the vacuum, the full
$SO(3,1)$ gauge symmetry is broken down to its the maximal compact
subgroup $SO(3)$. The ten scalars transform as
$\mathbf{1}+\mathbf{1}+\mathbf{3}+\mathbf{5}$ with the first singlet
being the dilaton.

\subsection{$AdS_7$ critical points}
We now investigate the vacuum structure of the $N=2$ gauged
supergravity with $SO(3,1)$ gauge group. We simplify the task by
restricting the potential to the two $SO(3)\subset SO(3,1)$ singlet
scalars. This truncation is consistent in the sense that all critical points found on this restricted scalar manifold are automatically critical points of the potential computed on the full scalar manifold as pointed out in \cite{warner}.
\\
\indent The scalar potential on these $SO(3)$ singlets is given by
\begin{eqnarray}
V&=&\frac{1}{16}e^{-\sigma-6\phi}\left[\left(1+8e^{2\phi}+3e^{4\phi}-32e^{6\phi}+3e^{8\phi}+8e^{10\phi}
+e^{12\phi}\right)g^2\right.\nonumber \\
& &\left.-32e^{\frac{5}{2}\sigma+3\phi}\left(1+e^{2\phi}+e^{4\phi}+e^{6\phi}\right)g h
+256h^2e^{5\sigma+6\phi}\right].\label{SO3_1_V1}
\end{eqnarray}
The scalar $\phi$ is an $SO(3)$ singlet coming from $SO(3,3)/SO(3)\times SO(3)$. It can be easily checked that this potential admits two critical points at $\phi=0$ and
\begin{equation}
\sigma=\frac{2}{5}\ln\frac{g}{16h},\qquad \textrm{and}\qquad \sigma=\frac{2}{5}\ln \frac{g}{8h}\, .
\end{equation}
As in the $SO(4)$ gauging studied in \cite{7D_flow}, the second
critical point is non-supersymmetric as can be checked by computing
the supersymmetry transformations of fermions. We will shift the
dilaton field so that the supersymmetric $AdS_7$ occurs at
$\sigma=0$. This is effectively achieved by setting $g=16h$. The
gauge group $SO(3,1)$ is broken down to its maximal compact subgroup
$SO(3)$, so the two critical points have $SO(3)$ symmetry. At these
critical points, the values of the cosmological constant ($V_0$) and
the $AdS_7$ radius $(L)$ are given in Table \ref{table13}.
\begin{table}[h]
\centering
\begin{tabular}{|c|c|c|c|}
  \hline
  Critical point & $\sigma$ & $V_0$ & $L$ \\ \hline
  I & $0$ & $-240h^2$ & $\frac{1}{4h}$ \\
  II & $\frac{2}{5}\ln 2$ & $-160(2^{\frac{3}{5}})h^2$ & $\frac{\sqrt{3}}{2(2^\frac{4}{5})h}$ \\
  \hline
\end{tabular}
\caption{Supersymmetric and non-supersymmetric $AdS_7$ critical points in $SO(3,1)$ gauging}\label{table13}
\end{table}
\\
\indent In our convention, the relation between $V_0$ and $L$ is given by
$L=\sqrt{-\frac{15}{V_0}}$. We can compute scalar masses at the
trivial critical point, $\sigma=0$, as shown in the Table
\ref{table11}.
\begin{table}[h]
         \centering
\begin{tabular}{|c|c|c|}
  \hline
  $SO(3)_{\textrm{diag}}$ & $m^2L^2$ & $\Delta$ \\ \hline
  $\mathbf{1}$ & $-8$ & $4$ \\
  $\mathbf{1}$ & $40$ & $10$ \\
  $\mathbf{3}$ & $0$  & $6$ \\
  $\mathbf{5}$ & $16$ & $8$ \\
  \hline
\end{tabular}
\caption{Scalar masses at the supersymmetric $AdS_7$ critical point
in $SO(3,1)$ gauging} \label{table11}
     \end{table}

\begin{table}[h]
         \centering
\begin{tabular}{|c|c|c|}
  \hline
  $SO(3)$ & $m^2L^2$ & $\Delta$ \\ \hline
  $\mathbf{1}$ & $12$ & $3+\sqrt{21}$ \\
  $\mathbf{1}$ & $36$ & $3(1+\sqrt{5})$ \\
  $\mathbf{3}$ & $0$  & $6$ \\
  $\mathbf{5}$ & $0$ & $6$ \\
  \hline
\end{tabular}
 \caption{Scalar masses at the non-supersymmetric $AdS_7$ critical point in $SO(3,1)$ gauging}\label{table12}
     \end{table}
In the table, we have given the representations under the unbroken
$SO(3)\subset SO(3,1)$ symmetry. The conformal dimension $\Delta$ of
the dual operators in the six-dimensional SCFT is also given. The
three scalars in the $\mathbf{3}$ representation correspondence to
the Goldstone bosons in the symmetry breaking $SO(3,1)$ to $SO(3)$.
These scalars correspond to marginal operators of dimension six.
From the table, we see that only the operator dual to the dilaton is
relevant. The other are either marginal or irrelevant.
\\
\indent Unlike in the $SO(4)$ gauging in which the
non-supersymmetric $AdS_7$ is unstable, we find that, in $SO(3,1)$
gauging, it is indeed stable as can be seen from the scalar masses
given in Table \ref{table12}. From the table, we see that the operator dual to $\sigma$ becomes
irrelevant at this critical point. We then expect that there should
be an RG flow driven by this operator from the $N=2$ supersymmetric
fixed point to this CFT. The gravity solution would involve the
metric $g_{\mu\nu}$ and $\sigma$. Since the flow is
non-supersymmetric, the flow solution has to be found by solving the
full second-order field equations. In general, these equations do
not admit an analytic solution. We will not go into the detail of
this flow here and will not give the corresponding numerical flow
solution. A similar study in the case of pure $N=2$ $SU(2)$ gauged
supergravity can be found in \cite{non_SUSY7Dflow}.

\subsection{$AdS_5$ critical points}
We now look for a vacuum solution of the form $AdS_5\times S^2$. In
this case, an abelian gauge field is turned on. There are six gauge
fields $A^I$, $I=1,\ldots, 6$, of $SO(3,1)$ in which the first three
gauge fields are those of the compact subgroup $SO(3)$. We will
choose the non-zero gauge field to be $A^3$. The seven-dimensional
metric is given by
\begin{equation}
ds^2=e^{2f(r)}dx_{1,3}^2+dr^2+e^{2g(r)}(d\theta^2+\sin^2d\phi^2)\label{AdS5S2_metric}
\end{equation}
where $dx^2_{1,3}$ is the flat metric on the four-dimensional
Minkowski space. The ansatz for the gauge field is given by
\begin{equation}
A^3=a\cos \theta d\phi, \qquad F^3=-a\sin \theta d\theta\wedge d\phi\, .
\end{equation}
From the metric, we can compute the following spin connections
\begin{eqnarray}
\omega^{\hat{\phi}}_{\phantom{s}\hat{\theta}} &=&e^{-g(r)}\cot\theta e^{\hat{\phi}},\qquad
\omega^{\hat{\phi}}_{\phantom{s}\hat{r}}=g(r)'e^{\hat{\phi}},\nonumber \\
\omega^{\hat{\theta}}_{\phantom{s}\hat{r}}&=&g(r)'e^{\hat{\theta}},\qquad \omega^{\hat{\mu}}_{\phantom{s}\hat{r}}=f'e^{\hat{\mu}}\, .
\end{eqnarray}
\indent From $SO(3,3)/SO(3)\times SO(3)$ coset, there are three
singlets under this $SO(2)\subset SO(3)$. One of them is the $SO(3)$
singlet mentioned before. The other two come from $\mathbf{3}$ and
$\mathbf{5}$ representations of $SO(3)$ with the former being one of
the three Goldstone bosons. We can then set up relevant BPS
equations by computing the supersymmetry transformations of
$\psi_\mu$, $\chi$ and $\lambda^r$. We will not give $\delta
\psi_r=0$ equation here. This will give rise to the equation for the
Killing spinors as a function of $r$.
\\
\indent We then impose the projections
\begin{equation}
\gamma_r\epsilon=\epsilon\qquad \textrm{and} \qquad i\gamma^{\hat{\theta}\hat{\phi}}\sigma^3\epsilon=\epsilon
\end{equation}
where hatted indices are tangent space indices. By imposing the twist condition
\begin{equation}
ag=1,
\end{equation}
we find that equation $\delta \psi_\theta=0$ is the same as
$\delta\psi_\phi=0$. The Killing spinors are then given by constant
spinors on $S^2$. Equations $\delta\psi_\mu$, $\mu=0,1,2,3$ lead to
a single equation for $f(r)$. With all these, we find the following
set of the BPS equations
\begin{eqnarray}
\phi_1'&=&\frac{e^{-\frac{\sigma}{2}-2\phi_1+2\phi_2-\phi_3}\left(1+e^{2\phi_3}\right)
\left(e^{2\phi_3}-1\right)g}{2\left(1+e^{4\phi_2}\right)},\\
\phi_2'&=&0,\label{phi2_eq}\\
\phi_3'&=&-\frac{1}{4}e^{-\frac{\sigma}{2}-2\phi_1-\phi_3-2g(r)}\left[2ae^{\sigma+2\phi_1}\left(e^{2\phi_3}-1\right)
\right. \nonumber \\
& &\left. -e^{2g(r)}\left(2e^{2\phi_1}+e^{4\phi_1}-e^{2\phi_3}-2e^{2(\phi_1+\phi_3)}+e^{4\phi_1+2\phi_3}-1\right)g\right],\\
\sigma'&=&\frac{1}{10}e^{-\frac{\sigma}{2}-2\phi_1-\phi_3-2g(r)}\left[2ae^{\sigma+2\phi_1}\left(1+e^{2\phi_3}
\right)+64he^{\frac{5}{2}\sigma+2\phi_1+\phi_3+2g(r)}\right.\nonumber \\
&
&\left.\phantom{e^{\frac{1}{2}}}-e^{2g(r)}\left(1-2e^{2\phi_1}-e^{4\phi_1}-e^{2\phi_3}-2e^{2(\phi_1+\phi_3)}+e^{4\phi_1+2\phi_3}\right)
g\right],\\
g(r)'&=&-\frac{2}{5}ae^{\frac{\sigma}{2}-\phi_3-2g(r)}\left(1+e^{2\phi_3}\right)+\frac{4}{5}he^{2\sigma}\nonumber \\
& &+\frac{1}{20}e^{-\frac{\sigma}{2}-2\phi_1-\phi_3}\left(1-2e^{2\phi_1}-e^{4\phi_1}-e^{2\phi_3}-2e^{2(\phi_1+\phi_3)}
+e^{4\phi_1+2\phi_3}\right)g,\quad\\
f'&=&\frac{1}{10}ae^{\frac{\sigma}{2}-\phi_3-2g(r)}\left(1+e^{2\phi_3}\right)+\frac{4}{5}he^{2\sigma}\nonumber \\
& &+\frac{1}{20}e^{-\frac{\sigma}{2}-2\phi_1-\phi_3}\left(1-2e^{2\phi_1}-e^{4\phi_1}-e^{2\phi_3}-2e^{2(\phi_1+\phi_3)}
+e^{4\phi_1+2\phi_3}\right)g\quad
\end{eqnarray}
where $\phi_i$, $i=1,2,3$ are the three singlets from
$SO(3,3)/SO(3)\times SO(3)$. The $'$ denotes $\frac{d}{dr}$. To
avoid the confusion with the gauge coupling $g$, we have explicitly
written the $S^2$ warp factor as $g(r)$.
\\
\indent $\phi_2$, being one of the Goldstone bosons, disappears
entirely from the scalar potential which, for these $SO(2)$
singlets, is given by
\begin{eqnarray}
V&=&\frac{1}{16}e^{-\sigma-4\phi_1-2\phi_3}\left[\left(1+2e^{4\phi_1}+e^{4\phi_3}+2e^{4(\phi_1+\phi_3)}
-16e^{4\phi_1+2\phi_3}+e^{8\phi_1+4\phi_3}\right)g^2\right.\nonumber \\
&
&+32ghe^{\frac{5\sigma}{2}+2\phi_1+\phi_3}\left(1-2e^{2\phi_1}-e^{4\phi_1}
-e^{2\phi_3}-2e^{2(\phi_1+\phi_3)}+e^{4\phi_1+2\phi_3}\right)\nonumber
\\
& &\left.+256h^2e^{5\sigma+4\phi_1+2\phi_3}\right].
\end{eqnarray}
When $\phi_3=\phi_1$, this reduces to the $SO(3)$ invariant
potential \eqref{SO3_1_V1}. Equation \eqref{phi2_eq} implies that $\phi_2$ is a constant. We will choose $\phi_2=0$ from now on in order to be consistent with the supersymmetric $AdS_7$ critical point.
\\
\indent The $AdS_5\times S^2$ geometry is characterized by the fixed
point solution of $g(r)'=\phi_i'=\sigma'=0$. From the above
equations, there is a solution only for $\phi_i=0$ and
\begin{eqnarray}
\sigma=\frac{2}{5}\ln \frac{g}{12h},\qquad g(r)=-\frac{1}{2}\ln\frac{g}{3a}+\frac{1}{5}\ln\frac{g}{12h}\, .
\end{eqnarray}
Near this fixed point with $g=16h$, we find $f\sim
\left(\frac{512}{9}\right)^{\frac{2}{5}}hr$. Therefore, the $AdS_5$
radius is given by
$L_{AdS_5}=\frac{1}{h}\left(\frac{9}{512}\right)^{\frac{2}{5}}$. At
this fixed point, the projection $\gamma_r\epsilon=\epsilon$ is not
needed, so the number of unbroken supercharges is eight. According
to the AdS/CFT correspondence, we will identify this $AdS_5$
solution with an $N=1$ SCFT in four dimensions.

\subsection{RG flows from 6D $N=(1,0)$ SCFT to 4D $N=1$ SCFT}
\begin{figure}
         \centering
         \begin{subfigure}[b]{0.4\textwidth}
                 \includegraphics[width=\textwidth]{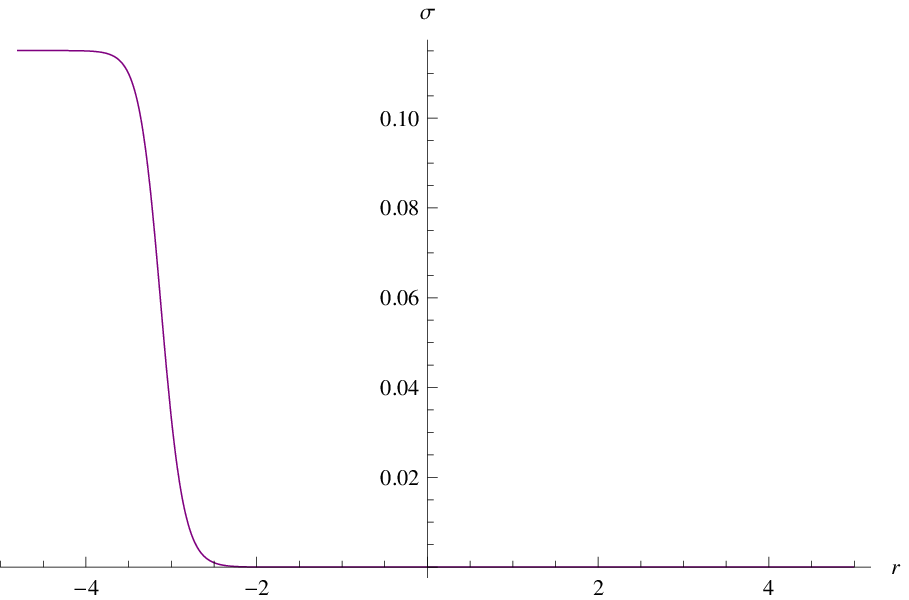}
                 \caption{A solution for $\sigma$}
         \end{subfigure}%
         ~ 
           \quad
         \begin{subfigure}[b]{0.4\textwidth}
                 \includegraphics[width=\textwidth]{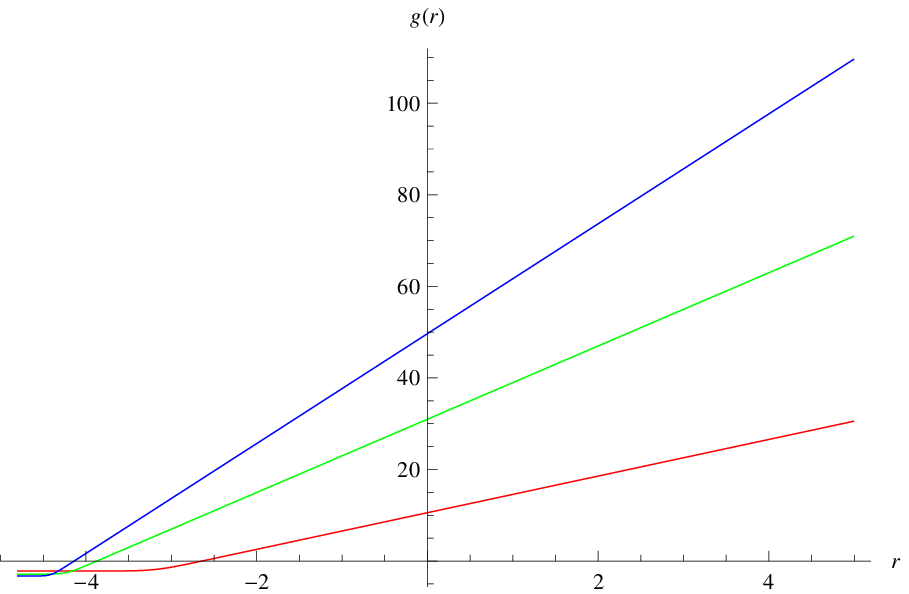}
                 \caption{Solutions for $g(r)$}
         \end{subfigure}
         \caption{RG flow solutions from $N=(1,0)$ SCFT in six dimensions to four-dimensional
         $N=1$ SCFT with the $g(r)$ solution given for three different values of $a$; $a=1$ (red),$a=2$ (green),$a=3$ (blue)}\label{fig1}
 \end{figure}

The existence of $AdS_5\times S^2$ geometry indicates that the
$N=(1,0)$ SCFT in six dimensions corresponding to $AdS_7$ critical
point can undergo an RG flow to a four-dimensional $N=1$ SCFT. We
begin the study of this RG flow solution by rewriting
the BPS equations for $\phi_i=0$
\begin{eqnarray}
\sigma'&=&\frac{2}{5}e^{-\frac{\sigma}{2}}\left(ae^{\sigma-2g(r)}+g-16he^{\frac{5\sigma}{2}}\right),\\
g(r)'&=&\frac{1}{5}e^{-\frac{\sigma}{2}}\left(g-4ae^{\sigma-2g(r)}+4he^{\frac{5\sigma}{2}}\right),\\
f'&=&\frac{1}{5}e^{-\frac{\sigma}{2}}\left(g+ae^{\sigma-2g(r)}+4he^{\frac{5\sigma}{2}}\right).
\end{eqnarray}
Near the IR $AdS_5$ fixed point, we find
\begin{eqnarray}
& &\sigma\sim g(r)\sim e^{(\sqrt{7}-1)\frac{r}{L_{AdS_5}}},\nonumber\\
& &f\sim \frac{r}{L_{AdS_5}}\, .
\end{eqnarray}
We then conclude that the operators dual to $\sigma$ and $g(r)$
become irrelevant in four dimensions with dimension
$\Delta=3+\sqrt{7}$. We are not able to find an analytic solution to
the above equations. We therefore give an example of numerical
solutions in Figure \ref{fig1}.
\\
\indent At the IR fixed point, the value of $\sigma$ does not depend
on $a$, but different values of $a$ give rise to different solutions for
$g(r)$. In Figure \ref{fig1}, we have given some examples of the $g(r)$
solutions with three different values of $a$, $a=1,2,3$ with $g=16h$ and
$h=1$. From the solutions, we see that, at large $r$, $g(r)\sim r$
and $\sigma\sim 0$. Furthermore, as $g(r)\sim r\rightarrow \infty$,
we find $f(r)\sim g(r)\sim r$. The UV geometry is $AdS_7$
corresponding to the six-dimensional $N=(1,0)$ SCFT. The behavior of
$\sigma$ near the UV point is given by
\begin{equation}
\sigma\sim e^{-\frac{4r}{L_{AdS_7}}}
\end{equation}
which indicates that the flow is driven by a VEV of a dimension-four
operator.

\subsection{$AdS_5\times H^2$ geometry}
We now consider a fixed point of the form $AdS_5\times H^2$ with
$H^2$ being a genus $g>1$ Riemann surface. In this case, we take the
metric ansatz to be
\begin{equation}
ds^2=e^{2f(r)}dx^2_{1,3}+dr^2+\frac{e^{2g(r)}}{y^2}(dx^2+dy^2).\label{metric_AdS5_H2}
\end{equation}
The $SO(2)$ gauge field is then given by
\begin{equation}
A=\frac{a}{y}dx,\qquad F=\frac{a}{y^2}dx\wedge dy\, .
\end{equation}
The spin connections computed from the above metric are given by
\begin{equation}
\omega^{\hat{x}}_{\phantom{s}\hat{r}}=g(r)'e^{\hat{x}},\qquad
\omega^{\hat{y}}_{\phantom{s}\hat{r}}=g(r)'e^{\hat{y}},\qquad
\omega^{\hat{x}}_{\phantom{s}\hat{y}}=-e^{-g(r)}e^{\hat{x}}\, .
\end{equation}
The twisted condition is still given by $ga=1$. The BPS equations
change by some signs, and it is still true that the $AdS_5$ is
possible only for $\phi_i=0$. The BPS equations, for $\phi_i=0$, are
then given by
\begin{eqnarray}
\sigma'&=&\frac{2}{5}e^{-\frac{\sigma}{2}}\left(-ae^{\sigma-2g(r)}+g-16he^{\frac{5\sigma}{2}}\right),\\
g(r)'&=&\frac{1}{5}e^{-\frac{\sigma}{2}}\left(g+4ae^{\sigma-2g(r)}+4he^{\frac{5\sigma}{2}}\right),\\
f'&=&\frac{1}{5}e^{-\frac{\sigma}{2}}\left(g-ae^{\sigma-2g(r)}+4he^{\frac{5\sigma}{2}}\right).
\end{eqnarray}
The fixed point conditions $\sigma'=g(r)'=0$ have the solution
\begin{eqnarray}
\sigma=\frac{2}{5}\ln \frac{g}{12h},\qquad
g(r)=-\frac{1}{2}\ln\left[-\frac{g}{3a}\right]+\frac{1}{5}\ln\frac{g}{12h}\,
.
\end{eqnarray}
\indent In this case, there is no real solution for $g(r)$ since the
twisted condition requires that $g$ must have the same sign as $a$.
Therefore, we conclude that there is no supersymmetric $AdS_5\times
H^2$ solution for $SO(3,1)$ gauging.

\section{$SL(3,\mathbb{R})$ gauge group}\label{SL3}
In this section, we consider the $SL(3,\mathbb{R})$ gauge group. The
minimal scalar manifold to accommodate this eight-dimensional gauge
group is $SO(3,5)/SO(3)\times SO(5)$. The structure constants can be
obtained from the generators
$T_I=(i\lambda_2,i\lambda_5,i\lambda_7,\lambda_1,\lambda_3,\lambda_4,\lambda_6,\lambda_8)$
with $I=1,\ldots, 8$. $\lambda_i$ are the usual Gell-mann matrices.
\\
\indent Under $SL(3,\mathbb{R})$, the adjoint representation of $SO(3,5)$ decomposes as
\begin{displaymath}
\mathbf{28}\rightarrow \mathbf{8}+ \mathbf{10}+\mathbf{10}'\, .
\end{displaymath}
At the vacuum, the $SL(3,\mathbb{R})$ symmetry is broken down to
$SO(3)$ with the embedding $\mathbf{3}\rightarrow\mathbf{3}$.
Therefore, under $SO(3)$, the $\mathbf{28}$ of $SO(3,5)$ further
decomposes as
\begin{displaymath}
\mathbf{28}\rightarrow \mathbf{3}+\mathbf{5}+ \mathbf{3}+\mathbf{7}+\mathbf{3}+\mathbf{7}\, .
\end{displaymath}
The fifteen scalars transform under $SO(3)$ as
$\mathbf{3}+\mathbf{5}+\mathbf{7}$. The other representations
$\mathbf{3}+\mathbf{3}+\mathbf{7}$ combine into the adjoint
representation of the composite local $SO(3)\times SO(5)$ symmetry.

\subsection{$AdS_7$ critical points}
By computing the scalar potential, we find that there are two $AdS_7$
critical points with $SO(3)$ symmetry as in the $SO(3,1)$ gauging
for vanishing vector multiplet scalars. One of them is
supersymmetric, and the other one is non-supersymmetric. We will
similarly set $g=16h$ to bring the supersymmetric $AdS_7$ to
$\sigma=0$. The characteristics of these two critical points are the
same as in $SO(3,1)$ gauging, so we will not repeat them here.
However, scalar masses at these two critical point are different and
are given in Table \ref{table2_1} and \ref{table2_2}.
     \begin{table}
         \centering
        \begin{tabular}{|c|c|c|}
  \hline
  $SO(3)$ & $m^2L^2$ & $\Delta$ \\ \hline
  $\mathbf{1}$ & $-8$ & $4$ \\
  $\mathbf{3}$ & $112$ & $14$ \\
  $\mathbf{5}$ & $0$  & $6$ \\
  $\mathbf{7}$ & $72$ & $12$ \\
  \hline
\end{tabular}
         \caption{Scalar masses at the supersymmetric $AdS_7$ critical point in $SL(3,\mathbb{R})$ gauging}\label{table2_1}
     \end{table}
     \begin{table}
         \centering
        \begin{tabular}{|c|c|c|}
  \hline
  $SO(3)$ & $m^2L^2$ & $\Delta$ \\ \hline
  $\mathbf{1}$ & $12$ & $3+\sqrt{21}$ \\
  $\mathbf{3}$ & $96$  & $3+\sqrt{105}$ \\
  $\mathbf{5}$ & $0$ & $6$ \\
  $\mathbf{7}$ & $36$ & $3(1+\sqrt{5})$  \\
  \hline
\end{tabular}
         \caption{Scalar masses at the non-supersymmetric $AdS_7$ critical point in $SL(3,\mathbb{R})$ gauging}\label{table2_2}
     \end{table}
\\
\indent As in the previous case, the $SO(3)$ singlet is the dilaton.
In this case, there are five Goldstone bosons from the
$SL(3,\mathbb{R})\rightarrow SO(3)$ symmetry breaking. The
non-supersymmetric $AdS_7$ is stable as in the $SO(3,1)$ gauging and
can be interpreted as a unitary six-dimensional CFT. We then expect
that there should be an RG flow from the supersymmetric $AdS_7$ to
the non-supersymmetric one. As in the previous case, the flow is
driven by a VEV of the operator dual to the dilaton $\sigma$. In the
IR, the operator becomes irrelevant with dimension
$\Delta=3+\sqrt{21}$.

\subsection{$AdS_5$ critical points}
We now study possible $AdS_5$ fixed points. We will turn on a gauge
field of $SO(2)$ which is a subgroup of the compact subgroup
$SO(3)\subset SL(3,\mathbb{R})$. Among the fifteen scalars, there
are three singlets under this $SO(2)$, and we will denote them by
$\phi_i$, $i=1,2,3$. Each of the three $SO(3)$ representations,
$\mathbf{3}+\mathbf{5}+\mathbf{7}$, gives one $SO(2)$ singlet.
\\
\indent We again use the metric ansatz \eqref{AdS5S2_metric} and the
gauge field $A^3=a\cos\theta d\phi$. With the twisted condition
$ga=1$ and the projectors $\gamma_r\epsilon=\epsilon$ and
$i\gamma^{\hat{\theta}\hat{\phi}}\sigma^3\epsilon=\epsilon$, we
obtain a system of complicated BPS equations. Since these equations might be
useful for other applications, we explicitly give them here
\begin{eqnarray}
\phi_1' &=&\frac{\sqrt{3}ge^{-\frac{\sigma}{2}-2\phi_1-\frac{2}{\sqrt{3}\phi_3}}\left(e^{4\phi_1}-1\right)
\left(e^{4\phi_2}-1\right)\left(e^{\frac{4\phi_3}{\sqrt{3}}}-1\right)}{4\left(1+e^{4\phi_2}\right)},\\
\phi_2'&=&\frac{\sqrt{3}}{4}ge^{-\frac{\sigma}{2}-2\phi_2-\frac{2\phi_3}{\sqrt{3}}}\left(1+e^{4\phi_2}\right)
\left(e^{\frac{4\phi_3}{\sqrt{3}}}-1\right),\\
\phi_3'&=&\frac{1}{16}e^{-\frac{\sigma}{2}-2\phi_1-2\phi_2-\frac{2\phi_3}{\sqrt{3}}-2g(r)}
\left[4\sqrt{3}ae^{\sigma+2\phi_1+2\phi_2}\left(1-e^{\frac{4\phi_3}{\sqrt{3}}}\right)\right.\nonumber \\
& & +ge^{g(r)}\left(3e^{4\phi_1+4\phi_2+\frac{4\phi_3}{\sqrt{3}}}+3e^{4\phi_2+\frac{4\phi_3}{\sqrt{3}}}
-4\sqrt{3}e^{2\phi_1+2\phi_2+\frac{4\phi_3}{\sqrt{3}}}-3e^{4\phi_1+\frac{4\phi_3}{\sqrt{3}}}
-3e^{\frac{4\phi_3}{\sqrt{3}}}\right.\nonumber \\
& &\left.\left.+3e^{4(\phi_1+\phi_2)}+4\sqrt{3}e^{2(\phi_1+\phi_2)}+3e^{4\phi_2}-3e^{4\phi_1}
-3\right)\right],\\
\sigma' &=&\frac{1}{20}e^{-\frac{\sigma}{2}-2\phi_1-2\phi_2-\frac{2\phi_3}{\sqrt{3}}-2g(r)}\left[
4ae^{\sigma+2(\phi_1+\phi_2)}\left(1+e^{\frac{4\phi_3}{\sqrt{3}}}+128he^{\frac{5\sigma}{2}+2\phi_1+2\phi_2
+\frac{2\phi_3}{\sqrt{3}}+2g(r)}\right)\right.\nonumber \\
&
&ge^{2g(r)}\left(\sqrt{3}\left(1+e^{4\phi_1}\right)-\sqrt{3}e^{4\phi_2}-4e^{2(\phi_1+\phi_2)}-\sqrt{3}
e^{4(\phi_1+\phi_2)}-\sqrt{3}e^{\frac{4\phi_3}{\sqrt{3}}}\right.
\nonumber \\
& &\left.\left.-\sqrt{3}e^{4\phi_1+\frac{4\phi_3}{\sqrt{3}}}
-4e^{2\phi_1+2\phi_2+\frac{4\phi_3}{\sqrt{3}}}+\sqrt{3}e^{4\phi_2+\frac{4\phi_3}{\sqrt{3}}}+\sqrt{3}
e^{4\phi_1+4\phi_2+\frac{4\phi_3}{\sqrt{3}}}\right)\right],\\
g(r)'&=&-\frac{2}{5}ae^{\frac{\sigma}{2}-\frac{2\phi_3}{\sqrt{3}}-2g(r)}\left(1+e^{\frac{4\phi_3}{\sqrt{3}}}\right)
+\frac{4}{5}he^{2\sigma}\nonumber \\
& &-\frac{1}{40}ge^{-\frac{\sigma}{2}-2\phi_1-2\phi_2-\frac{2\phi_3}{\sqrt{3}}}\left[\sqrt{3}
\left(1+e^{4\phi_1}\right)-\sqrt{3}e^{4\phi_2}-4e^{2(\phi_1+\phi_2)}-\sqrt{3}e^{4(\phi_1+\phi_2)}
\right.\nonumber \\
& &\left.-\sqrt{3}e^{\frac{4\phi_3}{\sqrt{3}}}\left(1+e^{4\phi_1}\right) -4e^{2\phi_1+2\phi_2+\frac{4\phi_3}{\sqrt{3}}}+\sqrt{3}e^{4\phi_2+\frac{4\phi_3}{\sqrt{3}}}
+\sqrt{3}e^{4\phi_1+4\phi_2+\frac{4\phi_3}{\sqrt{3}}}\right],\\
f'&=&\frac{1}{10}ae^{\frac{\sigma}{2}-\frac{2\phi_3}{\sqrt{3}}-2g(r)}\left(1+e^{\frac{4\phi_3}{\sqrt{3}}}\right)
+\frac{4}{5}he^{2\sigma}\nonumber \\
& &-\frac{1}{40}ge^{-\frac{\sigma}{2}-2\phi_1-2\phi_2-\frac{2\phi_3}{\sqrt{3}}}\left[\sqrt{3}
\left(1+e^{4\phi_1}\right)-\sqrt{3}e^{4\phi_2}-4e^{2(\phi_1+\phi_2)}-\sqrt{3}e^{4(\phi_1+\phi_2)}
\right.\nonumber \\
& &\left.-\sqrt{3}e^{\frac{4\phi_3}{\sqrt{3}}}\left(1+e^{4\phi_1}\right) -4e^{2\phi_1+2\phi_2+\frac{4\phi_3}{\sqrt{3}}}+\sqrt{3}e^{4\phi_2+\frac{4\phi_3}{\sqrt{3}}}
+\sqrt{3}e^{4\phi_1+4\phi_2+\frac{4\phi_3}{\sqrt{3}}}\right].
\end{eqnarray}
It can be easily verified that the first three equations have a
fixed point solution only when $\phi_i=0$ for all $i=1,2,3$. The
remaining equations then reduce to the same form as in the $SO(3,1)$
case. The RG flow solutions can also be studied in a similar manner,
and we will not repeat it here.
\\
\indent As a final remark, we note here that similar to the previous
case, it is not possible to have an $AdS_5\times H^2$ solution.

\section{$SO(2,2)$ gauge group}\label{SO2_2}
Unlike the previous two cases, this gauging does not admit a
maximally supersymmetric $AdS_7$. The vacuum is rather a
half-supersymmetric domain wall. This is not unexpected since the
minimal superconformal algebra in six dimensions has $SU(2)_R$
R-symmetry, but the vacuum of this gauging has only $SO(2)\times
SO(2)$ symmetry. The minimal scalar manifold for embedding this
gauge group is $SO(3,3)/SO(3)\times SO(3)$. The embedding of
$SO(2,2)$ in $SO(3,3)$ is given by the following structure constants
\begin{equation}
f_{IJ}^{\phantom{sss}K}=(g_1\epsilon_{\bar{i}\bar{j}\bar{l}}\eta^{\bar{k}\bar{l}}
,g_2\epsilon_{\bar{r}\bar{s}\bar{t}}\eta^{\bar{q}\bar{t}})
\end{equation}
with $\bar{i}=1,2,6$, $\bar{r}=3,4,5$, $\eta_{\bar{i}\bar{j}}=(-1,-1,1)$ and $\eta_{\bar{r}\bar{s}}=(-1,1,1)$.

\subsection{Domain wall solutions}
The vacuum of this gauging will have $SO(2)\times SO(2)$ symmetry.
Among the nine scalars from $SO(3,3)/SO(3)\times SO(3)$, there is
one $SO(2)\times SO(2)$ singlet which will be denoted by $\phi$. The
scalar potential for $SO(2)\times SO(2)$ singlet scalars is given by
\begin{equation}
V=\frac{1}{2}g_1e^{-\sigma}+4g_1he^{\frac{3\sigma}{2}}\left(e^{-\phi}-e^\phi\right)+16h^2e^{4\sigma}\, .
\end{equation}
It can be checked that this potential does not admit any critical points unless $h=g_1=0$. The vacuum is then a domain wall.
\\
\indent To study the domain wall solution, we write down the associated BPS equations by setting all the fields but the metric and scalars to zero. The metric is given by the domain wall ansatz
\begin{equation}
ds^2=e^{2A(r)}dx_{1,5}^2+dr^2\, .
\end{equation}
With the projection $\gamma_r\epsilon=\epsilon$, the relevant BPS
equations read
\begin{eqnarray}
\phi'&=&-\frac{1}{2}g_1e^{-\frac{\sigma}{2}-\phi}\left(1+e^{2\phi}\right),\label{SO22_DW1}\\
\sigma'&=&\frac{1}{5}e^{-\frac{\sigma}{2}-\phi}\left[g_1\left(e^{2\phi}-1\right)
-32he^{\frac{5\sigma}{2}+\phi}\right]\label{SO22_DW2},\\
A'&=&\frac{1}{10}e^{-\frac{\sigma}{2}-\phi}\left[g_1\left(e^{2\phi}-1\right)
+8he^{\frac{5\sigma}{2}+\phi}\right].\label{SO22_DW3}
\end{eqnarray}
By changing the radial coordinate from $r$ to $\tilde{r}$ with the relation $\frac{d\tilde{r}}{dr}=e^{-\frac{\sigma}{2}}$, it is not difficult to find the solutions for $\phi$, $\sigma$ and $A$. These are given by
\begin{eqnarray}
\phi&=&\ln \left[\tan\frac{C_1-g_1\tilde{r}}{2}\right],\label{SO22phi_sol}\\
\sigma &=&\frac{2}{5}\phi-\frac{2}{5}\ln\left[ \frac{16h}{g_1}\left(4C_2(1+e^{2\phi})-1\right)\right],\label{SO22sigma_sol}\\
A&=&\frac{1}{5}\phi-\frac{1}{4}\ln (1+e^{2\phi})+\frac{1}{20}\ln
\left[1-4C_2\left(1+e^{2\phi}\right)\right]\label{SO22A_sol}
\end{eqnarray}
where $C_1$ and $C_2$ are integration constants. We have omitted the
additive constant to $A$ since this can be removed by rescaling
$dx_{1,5}^2$ coordinates. According to the general DW/QFT
correspondence, this solution should be dual to a non-conformal
$N=(1,0)$ gauge theory in six dimensions. As $\tilde{r}\rightarrow
\frac{C_1}{g_1}$, the two scalars are logarithmically divergent.
After changing the coordinate from $\tilde{r}$ back to $r$, we find
the behavior of $\phi$ and $\sigma$ as $\tilde{r}\sim
\frac{C_1}{g_1}$, which is equivalent to $r\sim \frac{C}{g_1}$,
\begin{equation}
\phi\sim \frac{5}{6}\ln\left[\frac{C-g_1r}{2}\right],\qquad
\sigma\sim
\frac{1}{3}\ln\left[\frac{C-g_1r}{2}\right]\label{DW_SO22}
\end{equation}
where $C$ is a new integration constant coming from solving for $\tilde{r}$ in term of $r$. After rescaling $dx^2_{1,5}$ coordinates, the metric in this limit is given by
\begin{equation}
ds^2=(C-g_1r)^\frac{1}{3}dx^2_{1,5}+dr^2\, .
\end{equation}

\subsection{$AdS_5$ critical points}
We now look for a vacuum solution of the form $AdS_5\times S^2$. In
this case, there are two abelian $SO(2)$ gauge groups. The
corresponding gauge fields are denoted by
\begin{equation}
A^3=a\sin\theta d\phi,\qquad A^6=b\sin\theta d\phi\, .
\end{equation}
The metric is still given by \eqref{AdS5S2_metric}. In order to find the BPS equations, we impose the projectors $\gamma_r\epsilon =\epsilon$ and $i\gamma^{\hat{\theta}\hat{\phi}}\sigma^3\epsilon=\epsilon$. The twisted condition is now given by
\begin{equation}
g_1b=1\, .
\end{equation}
\indent Proceed as in the previous cases but with one more gauge
field, we find the following BPS equations
\begin{eqnarray}
\phi'&=&\frac{1}{2}e^{-\frac{\sigma}{2}-\phi-2g(r)}\left[ae^\sigma\left(1-e^{2\phi}\right)-
\left(1+e^{2\phi}\right)\left(be^\sigma+e^{2g(r)}g_1\right)\right]\label{Eq1}, \\
\sigma'&=&\frac{1}{5}e^{-\frac{\sigma}{2}-\phi-2g(r)}\left[(a-b)e^\sigma+(a+b)e^{\sigma+2\phi}\right.\nonumber
\\
& &\left.+e^{2g(r)}
\left[\left(e^{2\phi}-1\right)g_1-32he^{\frac{5\sigma}{2}+\phi}\right]\right],\quad\label{Eq2}\\
g(r)'&=&\frac{1}{10}e^{-\frac{\sigma}{2}-\phi-2g(r)}\left[e^{2g(r)}\left[\left(e^{2\phi}-1\right)g_1
+8he^{\frac{5\sigma}{2}+\phi}\right]\right. \nonumber \\
& &\left.+4(b-a)e^\sigma-4(a+b)e^{\sigma+2\phi}\right]\label{Eq3},\\
f'&=&\frac{1}{10}e^{-\frac{\sigma}{2}-\phi-2g(r)}\left[e^{2g(r)}\left[\left(e^{2\phi}-1\right)g_1
+8he^{\frac{5\sigma}{2}+\phi}\right]\right.\nonumber
\\
& &\left.+(a-b)e^\sigma+(a+b)e^{\sigma+2\phi}\right]\label{Eq4}
\end{eqnarray}
where $\phi$ is the $SO(2)\times SO(2)$ singlet scalar from
$SO(3,3)/SO(3)\times SO(3)$.
\\
\indent The equations $\phi'=\sigma'=g(r)'=0$ admit a fixed point
solution given by
\begin{eqnarray}
\phi &=&\frac{1}{2}\ln \left[\frac{\sqrt{4b^2-3a^2}-a}{2(a+b)}\right],\nonumber \\
\sigma &=&\frac{1}{5}\ln \left[\frac{a^2g_1^2\left(\sqrt{4b^2-3a^2}-a\right)}{32(a+b)h^2
\left(2b-3a+\sqrt{4b^2-3a^2}\right)}\right],\nonumber \\
g(r)&=&\frac{1}{10}\ln\left[\frac{(a+b)^4\left(a-2b+\sqrt{4b^2-3a^2}\right)^5\left(3a-2b-\sqrt{4b^2-3a^2}\right)^3}
{1024a^3g_1^3h^2\left(a-\sqrt{4b^2-3a^2}\right)^4}\right].\qquad
\end{eqnarray}
\indent It can be checked that the solution exists for $g_1<0$ and
$a<0$ with $b>-a$ or $g_1<0$ with $a>0$ and $b>a$. This in turn
implies that $g_1$ and $b$ always have opposite sign in
contradiction with the twisted condition $g_1b=1$. Therefore, the
$SO(2,2)$ gauging does not admit $AdS_5\times S^2$ geometry.
\\
\indent However, there exists an $AdS_5\times H^2$ geometry. In this
case, we have the metric \eqref{metric_AdS5_H2} with the gauge
fields given by
\begin{equation}
A^3=\frac{a}{y}dx,\qquad A^6=\frac{b}{y}dx\, .
\end{equation}
The twisted condition is still given by $g_1b=1$. The BPS equations
are given by \eqref{Eq1}, \eqref{Eq2}, \eqref{Eq3} and \eqref{Eq4}
but with $(a,b)$ replaced by $(-a,-b)$. The values of scalar fields
at the $AdS_5$ fixed point solution are real for $g_1<0$ and $a<0$
with $b<a$ in compatible with the twisted condition. Furthermore, it
is not possible to have an $AdS_5$ fixed point with $a=\pm b$. This
rules out the possibility of $AdS_5$ fixed point with
$SO(2)_{\textrm{diag}}\subset SO(2)\times SO(2)$ symmetry. For
$a=0$, only one $SO(2)$ gauge field turned on, it can also be
checked that the $AdS_5$ fixed point does not exist. The $b=0$ case
is not possible since this is not consistent with the twisted
condition with finite $g_1$.

\subsection{RG flows from $N=1$ 4D SCFT to 6D $N=(1,0)$ SYM}
\begin{figure}
         \centering
         \begin{subfigure}[b]{0.3\textwidth}
                 \includegraphics[width=\textwidth]{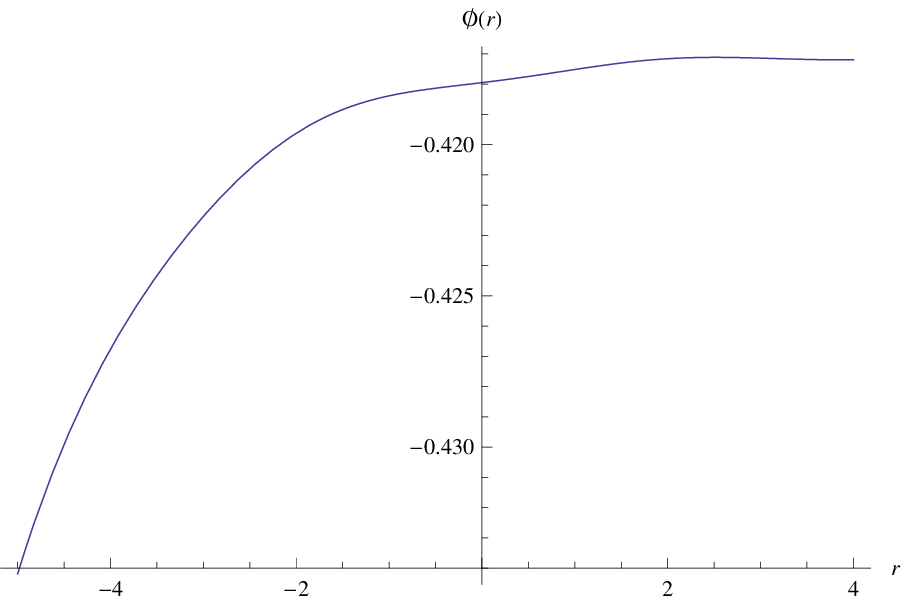}
                 \caption{A solution for $\phi$}
         \end{subfigure}%
         ~ 
         \begin{subfigure}[b]{0.3\textwidth}
                 \includegraphics[width=\textwidth]{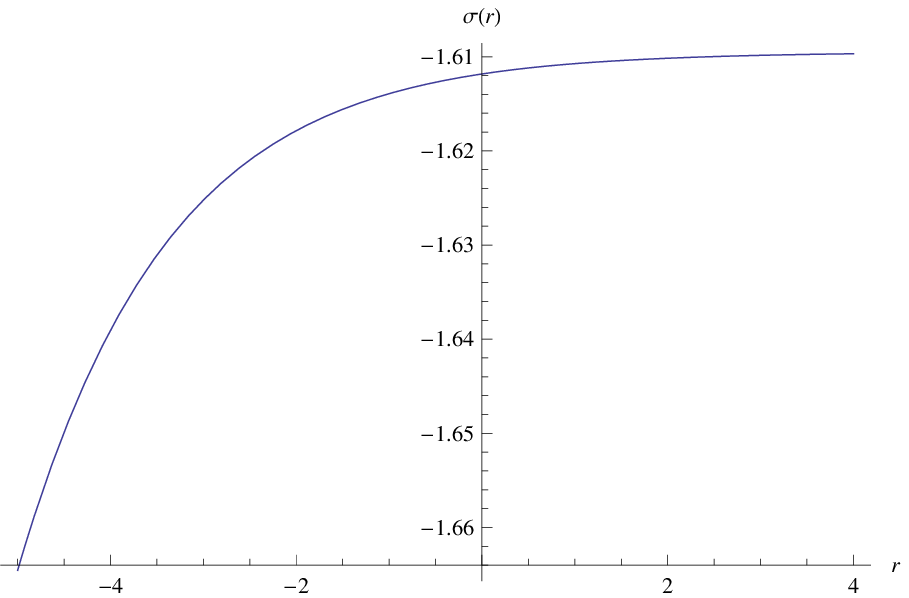}
                 \caption{A solution for $\sigma$}
         \end{subfigure}
         \begin{subfigure}[b]{0.3\textwidth}
                 \includegraphics[width=\textwidth]{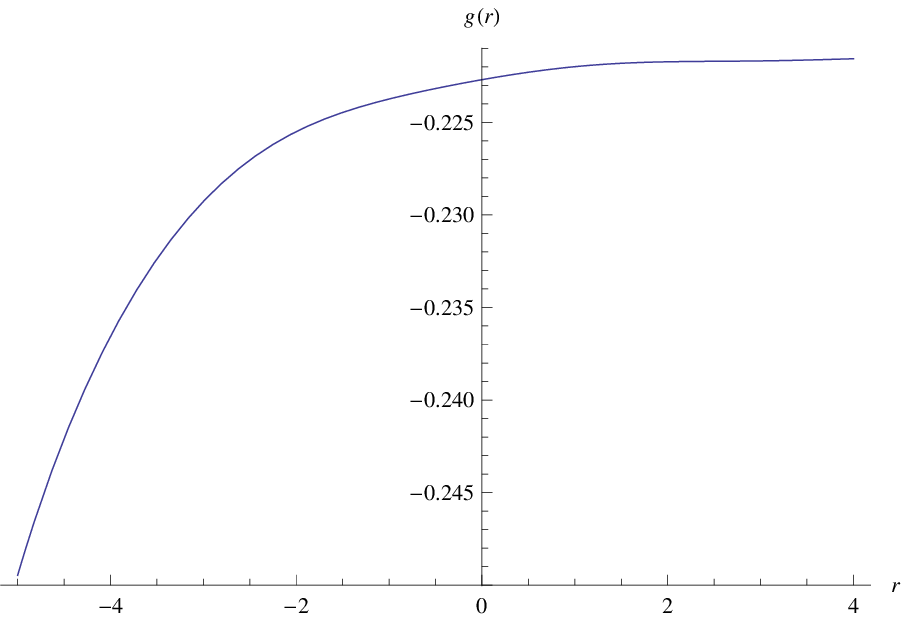}
                 \caption{A solution for $g(r)$}
         \end{subfigure}
         \caption{An RG flow solution from $N=1$ SCFT in four dimensions to six-dimensional $N=(1,0)$ SYM}\label{fig2}
 \end{figure}

According to the AdS/CFT correspondence, the existence of $AdS_5$ fixed point implies a dual $N=1$ SCFT in four dimensions. Near this $AdS_5$ critical point, the linearized BPS equations give
\begin{equation}
\phi\sim\sigma\sim g(r)\sim e^{-\frac{4r}{L}}
\end{equation}
where $L$ is the $AdS_5$ radius. We see that the $AdS_5$ should
appear in the UV identified with $r\rightarrow \infty$. This UV SCFT
in four dimensions undergoes an RG flow to a six-dimensional
$N=(1,0)$ SYM corresponding to the domain wall solution given by
equations \eqref{SO22phi_sol}, \eqref{SO22sigma_sol} and
\eqref{SO22A_sol}. In the IR, the warped factors behave as $f(r)\sim
g(r)\sim \ln(C-g_1r)^{\frac{1}{3}}$ while the behavior of the scalars
$\sigma$ and $\phi$ is given in \eqref{DW_SO22}. The flow is then
driven by vacuum expectations value of marginal operators dual to
$\phi$, $\sigma$ and $g(r)$. We give an example of numerical flow
solutions to the BPS equations in Figure \ref{fig2}. This solution
is found for particular values of $a=-1$, $b=-2$, $g=-\frac{1}{2}$
and $h=1$ which give
\begin{equation}
\phi=-0.4171,\qquad \sigma =-1.6095,\qquad g(r)=-0.2214
\end{equation}
at the $AdS_5$ fixed point.
 \\
\indent As usual in flows to non-conformal field theories, the
domain wall geometry in the IR is singular. We have checked that the
domain wall solution given in equation \eqref{DW_SO22} gives rise to
a good singularity according to the criterion of
\cite{Gubser_singularity}. Given the behavior of $\sigma$ and $\phi$ in
\eqref{DW_SO22}, we find that the scalar potential is bounded above
$V\rightarrow -\infty$. Therefore, the IR domain wall corresponds to
a physical gauge theory in six dimensions.

\section{$SO(2,1)$ and $SO(2,2)\times SO(2,1)$ gauge groups}\label{SO2_2_1}
In this section, we consider the last two possible non-compact gauge
groups $SO(2,1)$ and $SO(2,2)\times SO(2,1)$. We will see that both
of them admit a vacuum solution in the form of a domain wall.

\subsection{Vacua of $SO(2,1)$ gauging}
In this case, the minimal scalar manifold is given by
$SO(3,1)/SO(3)$. There are three scalars in this manifold. The
structure constants of the $SO(2,1)$ gauge group can be chosen to be
\begin{equation}
f_{IJK}=(g\epsilon_{\bar{i}\bar{j}\bar{k}},0),\qquad \bar{i}=1,2,4\, .
\end{equation}
This corresponds to choosing the $SO(2,1)$ generators to be $(T_{41},T_{42},T_{12})$ from the $SO(3,1)$ generators $(T_{ij},T_{4i})$, $i,j=1,2,3$.
\\
\indent The scalar potential does not have any critical points.
Therefore, we expect that the vacuum is a domain wall. Using the
domain wall ansatz for the metric and the projector
$\gamma_r\epsilon=\epsilon$, we find the BPS equations for all of
the four scalars
\begin{eqnarray}
\phi_1'&=&-\frac{e^{-\frac{\sigma}{2}-\phi_1}\left(e^{2\phi_1}-1\right)\left(e^{2\phi_3}-1\right)g}
{2\left(1+e^{2\phi_3}\right)},\\
\phi_2'
&=&-\frac{e^{-\frac{\sigma}{2}-\phi_2}\left(e^{2\phi_2}-1\right)\left(e^{2\phi_3}-1\right)g}
{2\left(1+e^{2\phi_3}\right)},\\
\phi_3'&=&-\frac{1}{2}e^{-\frac{\sigma}{2}-\phi_3}\left(1+e^{2\phi_3}\right)g,\\
\sigma'&=&\frac{1}{20}e^{-\frac{\sigma}{2}-\phi_1-\phi_2-\phi_3}\left(1+e^{2\phi_1}\right)
\left(1+e^{2\phi_2}\right)\left(e^{2\phi_3-1}\right)g-\frac{32}{5}he^{2\sigma},\\
A'&=&\frac{1}{40}e^{-\frac{\sigma}{2}-\phi_1-\phi_2-\phi_3}\left(1+e^{2\phi_1}\right)
\left(1+e^{2\phi_2}\right)\left(e^{2\phi_3-1}\right)g+\frac{4}{5}he^{2\sigma}\, .
\end{eqnarray}
In these equations, $\phi_i$, $i=1,2,3$ are scalars in $SO(3,1)/SO(3)$.
\\
\indent It is difficult to find an exact solution with all scalars
non-vanishing. On the other hand, a numerical solution could be obtained by the same procedure as in the previous sections. Since analytic solutions might be more interesting, we consider only a domain wall solution
preserving $SO(2)\subset SO(2,1)$ symmetry. Among these $\phi_i$'s,
$\phi_3$ is an $SO(2)$ singlet. It turns out that on this scalar
submanifold the solution is the same as that given in
\eqref{SO22phi_sol}, \eqref{SO22sigma_sol} and \eqref{SO22A_sol}
with $\phi$ replaced by $\phi_3$.

\subsection{Vacua of $SO(2,2)\times SO(2,1)$ gauging}
The last gauge group to be considered is $SO(2,2)\times SO(2,1)\sim SO(2,1)\times SO(2,1)\times SO(2,1)$. The minimal scalar manifold in this case is $SO(3,6)/SO(3)\times SO(6)$ with the embedding of $SO(2,2)\times SO(2,1)$ in $SO(3,6)$ given by the following structure constants
\begin{equation}
f_{IJ}^{\phantom{ssd}K}=(g_1\epsilon_{\bar{i}\bar{j}\bar{k}}\eta^{\bar{k}\bar{l}},
g_2\epsilon_{\bar{r}\bar{s}\bar{t}}\eta^{\bar{t}\bar{q}},g_3\epsilon_{\tilde{i}\tilde{j}\tilde{k}}
\eta^{\tilde{k}\tilde{l}}),\qquad \bar{i}=1,4,5,\quad \bar{r}=2,6,7,\quad \tilde{i}=3,8,9\, .
\end{equation}
The Killing metrics are given by $\eta_{\bar{i}\bar{j}}=(-1,1,1)$, $\eta_{\bar{r}\bar{s}}=(-1,1,1)$ and $\eta_{\tilde{i}\tilde{j}}=(-1,1,1)$, and $g_1$, $g_2$ and $g_3$ are gauge couplings of the three $SO(2,1)$ factors.
\\
\indent Apart from the dilaton, there are no scalars which are singlet under the maximal compact subgroup $SO(2)\times SO(2)\times SO(2)$. However, it can be shown that the potential does not have any critical points for $g_i,h\neq 0$. A simple domain wall solution can be obtained by solving the BPS equations for $\sigma$ and the metric. There might be other solutions with non-vanishing scalars from $SO(3,6)/SO(3)\times SO(6)$, but we have not found any of them. Therefore, we will restrict ourselves to the domain wall with only $\sigma$ and the metric non-vanishing. Using the projector $\gamma_r\epsilon=\epsilon$ as usual, we find the following BPS equations
\begin{eqnarray}
\sigma'&=&-\frac{32}{5}e^{2\sigma}h,\\
A'&=&=\frac{4}{5}e^{2\sigma}h\, .
\end{eqnarray}
These equations can be readily solved for the solution
\begin{eqnarray}
\sigma &=&-\frac{1}{2}\ln \left[\frac{64hr}{5}+C\right],\\
A&=&\frac{1}{16}\ln \left[\frac{64hr}{5}+C\right]
\end{eqnarray}
where $C$ is an integration constant. The seven-dimensional metric is given by
\begin{equation}
ds^2=(64hr+5C)^{\frac{1}{8}}dx^2_{1,5}+dr^2
\end{equation}
 where we have rescaled the $dx^2_{1,5}$ coordinates by $\frac{1}{5}$.
 \\
 \indent For $h=0$, there is a Minkowski vacuum with $V_0=0$. All scalar masses at this critical point are given in Table \ref{Minkowski}. The $SO(2)^3$ singlet is the dilaton which is massless while the other six massless scalars are Goldstone bosons of the symmetry breaking $SO(2,1)^3\rightarrow SO(2)^3$.
\begin{table}
         \centering
        \begin{tabular}{|c|c|}
  \hline
  $m^2$ & $SO(2)\times SO(2)\times SO(2)$ representation \\ \hline
  $0$ & $(\mathbf{1},\mathbf{1},\mathbf{1})$  \\
  $0$ & $(\mathbf{1},\mathbf{1},\mathbf{2})+(\mathbf{1},\mathbf{2},\mathbf{1})
  +(\mathbf{2},\mathbf{1},\mathbf{1})$  \\
  $g_1^2$ & $2\times(\mathbf{2},\mathbf{1},\mathbf{1})$  \\
  $g_2^2$ &  $2\times(\mathbf{1},\mathbf{2},\mathbf{1})$  \\
  $g_3^2$ & $2\times(\mathbf{1},\mathbf{1},\mathbf{2})$ \\
  \hline
\end{tabular}
         \caption{Scalar masses at the supersymmetric Minkowski vacuum in $SO(2,2)\times SO(2,1)$ gauging}\label{Minkowski}
     \end{table}

\section{Conclusions}\label{conclusion}
We have studied $N=2$ gauged supergravity in seven dimensions with
non-compact gauge groups. In $SO(3,1)$ and $SL(3,\mathbb{R})$
gaugings, we have found new supersymmetric $AdS_7$ critical points.
These should correspond to new $N=(1,0)$ SCFTs in six dimensions. We
have also found that there exist $AdS_5\times S^2$ solutions to
these gaugings. The solutions preserve eight supercharges and should
be dual to some $N=1$ four-dimensional SCFT with $SO(2)\sim U(1)$
global symmetry identified with the R-symmetry. We have then studied
RG flows from the six-dimensional $N=(1,0)$ SCFT to the $N=1$ SCFT
in four dimensions and argued that the flow is driven by a vacuum
expectation value of a dimension-four operator dual to the
supergravity dilaton. A numerical solution for an example of these
flows has also been given. In addition, we have shown that both of
the gauge groups admit a stable non-supersymmetric $AdS_7$ solution
which should be interpreted as a unitary CFT. This is not the case
for the compact $SO(4)$ gauging studied in \cite{7D_flow} in which
the non-supersymmetric critical point has been shown to be unstable.
\\
\indent In the $SO(2,2)$ gauging, we have given a domain wall vacuum
solution preserving half of the supersymmetry. According to the
DW/QFT correspondence, this is expected to be dual to a
non-conformal SYM in six dimensions. This $SO(2,2)$ gauging does not
admit an $AdS_5\times S^2$ solution but an $AdS_5\times H^2$
geometry with eight supercharges. The latter corresponds to an $N=1$
SCFT in four dimensions with $SO(2)\times SO(2)$ global symmetry. It
is likely that the a-maximization \cite{a-max,a-max_Tajikawa,a-max_Phil} is needed in order to identify the
correct $U(1)_R$ symmetry out of the $SO(2)\times SO(2)$ symmetry. We
have studied an RG flow from this SCFT to a non-conformal SYM in six
dimensions, dual to the seven-dimensional domain wall, and argued
that the flow is driven by vacuum expectation values of marginal
operators. We have also investigated $SO(2,1)$ and $SO(2,2)\times
SO(2,1)$ gaugings. Both of them admit a half-supersymmetric domain
wall as a vacuum solution. For vanishing topological mass, the
$SO(2,2)\times SO(2,1)$ gauging admits a seven-dimensional Minkowski
vacuum preserving all of the supersymmetry and $SO(2)\times
SO(2)\times SO(2)$ symmetry.
\\
\indent Due to the existence of new supersymmetric $AdS_7$ critical
points, the results of this paper might be useful in AdS$_7$/CFT$_6$
correspondence within the framework of seven-dimensional gauged
supergravity. The new $AdS_5$ backgrounds could be of interest in
the context of AdS$_5$/CFT$_4$ correspondence. RG flows across
dimensions described by gravity solutions connecting these
geometries would provide additional examples of flows in twisted
field theories. It is also interesting, if possible, to identify
these $AdS_5$ critical points with the known four-dimensional SCFTs.
\\
\indent
Until now, only the embedding of the $SO(4)$ gauging of $N=2$ supergravity coupled to three vector multiplets in eleven-dimensional supergravity has been given \cite{SO4_7Dfrom11D}. The embedding of non-compact gauge groups in ten or eleven dimensions in the presence of topological mass term is presently not known. It would be of particular interest to find such an embedding so that the results reported here would be
given an interpretation in terms of brane configurations in string/M
theory.
\acknowledgments
This work is supported by Chulalongkorn University through Research Grant for New Scholar Ratchadaphiseksomphot Endowment Fund under grant RGN-2557-002-02-23.
The author is also supported by The Thailand Research Fund (TRF)
under grant TRG5680010.


\end{document}